\documentclass[12pt]{article}
\usepackage{amssymb}
\usepackage{amsmath}
\usepackage{amscd}
\usepackage{latexsym}
\usepackage{graphicx}
\usepackage{url}
\usepackage{epsfig} 
\usepackage{listings}
\usepackage{longtable}
\usepackage{hyperref}

\usepackage{xcolor}

\catcode`@=11
\makeatletter\renewcommand{\section}{\@startsection
{section}{1}{\z@}{-3.5ex plus -1ex minus
    -.2ex}{2.3ex plus .2ex}{\large\bf }}

\makeatletter\renewcommand{\subsection}{\@startsection{subsection}{2}{\z@}{-3.25ex
plus -1ex minus
   -.2ex}{1.5ex plus .2ex}{\bf }}

\numberwithin{equation}{section}
\newcounter{saveeqn}

\catcode`@=12
\def\bea{\begin{eqnarray}}
\def\eea{\end{eqnarray}}

\def\1{{\bar 1}}
\def\2{{\bar 2}}
\def\3{{\bar 3}}
\def\4{{\bar 4}}

\def\+{\dagger}
\def\={\ =\ }

\def\and{\quad\textrm{and}\quad}

\begin{document}

\begin{titlepage}
\setcounter{page}{0}

\vskip 2.0cm

\begin{center}

{\Large\bf
Computational lower limits on small Ramsey numbers
}

\vspace{12mm}

{\large Eugene Kuznetsov

nameless@fastmail.fm
}

\vspace{12mm}

\begin{abstract}
\noindent

Computer-based attempts to construct lower bounds for small Ramsey numbers are discussed. A systematic review of cyclic Ramsey graphs is attempted. Many known lower bounds are reproduced. Several new bounds are reported.

\end{abstract}

\end{center}
\end{titlepage}

\section{Introduction}

A 2-color coloring of a complete graph is a Ramsey (k,j) coloring if it contains no monochromatic 0-color cliques of order k and no monochromatic 1-color cliques of order j. 2-color Ramsey number R(k,j) is defined as the smallest number N for which no (k,j) colorings of order N exist. (Therefore, to establish a lower bound of $R(k,j)\geq x$, one can exhibit a (k,j) coloring of a graph of order x-1.)

For the purposes of this article, let's define a "distance 2-color coloring" of a complete N-graph, with vertices labeled with numbers 1 to N (henceforth, "N-coloring"), as an assignment of a color (represented as a bit 0 or 1) C(a,b) to each link in the graph in such a way that C(a,b)=C(a+k,b+k) for any $1 \leq a,b,a+k,b+k \leq N, a \neq b$. It may be conveniently represented by a computer as a bit mask with N-1 bits, with each bit corresponding to colors of link (1,2), (1,3), ... (1,N). 

Distance colorings are interesting because they include cyclic colorings as a subset (a cyclic, or circulant, coloring is a coloring that is invariant under the rotation of vertex labels), and cyclic colorings represent many largest known Ramsey graphs: for example, both the largest R(4,4) graph and the largest known R(6,6) graph are cyclic. (It's not hard to see that a distance coloring is a cyclic coloring iff its bit mask is symmetric: $b_k = b_{N-2-k}$ for $0 \leq k<N-1$.) However, cyclic colorings are difficult to enumerate directly. A convenient property of distance colorings is that any M-subset of the first M vertices of a N-coloring is itself a distance coloring, and, conversely, adding vertices to the end of a distance coloring, with proper assignments of newly created links, creates new distance colorings. This permits us to "grow" colorings, starting with a short coloring and recursively enumerating its possible extensions. In a search for large (k,j) graphs, we'd skip colorings with observed complete k/j-cliques, thus having to analyze fewer than $2^{N-1}$ colorings at each N.

Algorithms described in this paper fall into two main categories. For lower (k,j) we can do full enumeration and to find all Ramsey cyclic and distance colorings. With these algorithms and on present-day computer hardware, this is feasible up to R(3,21), R(4,13), R(5,8) and R(6,7). For higher (k,j), two complementary methods are discussed which permit us to locate many if not all cyclic colorings in reasonable amounts of time (though without making any guarantees that resulting lists are complete.) 

\section{Full enumeration}

It's convenient to divide the search space into parts corresponding to "signatures". An $s$-signature is a fully colored distance graph on $s$ vertices (represented as a S-1 bit mask.) For example, there are $2^{34} \approx 1.7*10^{10}$ 35-signatures ($8.6*10^9$ if we're working with a diagonal Ramsey number.) Any distance coloring with more than 35 verticles is a descendant of one of them. 

To enumerate graphs, depth-first search is optimal, since it is generally faster than breadth-first and its memory footprint is low, allowing us to keep all necessary lists of cliques in system memory and minimize copying. At each step of the process, we color a link, review the graph for newly created cliques (appendix A), continue if no problems were detected (no complete color-0 k-cliques or color-1 j-cliques), try the next color or backtrack otherwise.

To improve performance of the search, we define a parameter $d$, which specifies the minimum size of the colorings we're interested in, and, for colorings with less than '$d$' vertices, add the following operations (in essence, trying to look ahead and to eliminate dead-end colorings as early as possible.)

* Forced link search (appendix B). At each pass, review the graph for k/j-cliques which are short one link of completion. Finding such a clique forces us to color the missing link in the color that is opposite to the clique. At each step it is sufficient to consider only cliques containing the link that has just been colored. This in turn would possibly create 1-short k/j-cliques of the opposite color, etc. At larger N this process would frequently result in complete k-cliques, allowing us to exclude the entire set of descendants of this coloring from consideration.

* Forced link group search (appendix C). This involves looking for k-cliques which are short 2 (or, optionally, 3) links of completion (such a clique would mean a prohibition to color all links the same color as the clique.) This process is more costly in terms of complexity vs. benefit (whereas a 1-short clique cuts the search space in half, a 2-short clique only reduces it by 1/4 and a 3-short clique reduces it by 1/8), but it occasionally results in forced links as well, and can provide input for the next method:

* Out of order coloring. Rather than coloring all links in increasing order, we can pick links that cut the search space the most, by looking at the numbers of known link pairs (generated during the previous step) associated with each potential link. 

* Graph rebuild. This is needed, because, unless links are colored in increasing order, the algorithm in appendix A misses some cliques. For example, if we have a coloring with links 10, 60 and 70 set to 1 and link 50 unset, setting 50 to 1 creates a clique (0,10,60,70). Detecting such cliques is expensive, and it's preferable to do an ordered rebuild once in a while instead.

The first of these four operations is performed each time a link is colored, and then recursively until no more changes can be detected. The other three are relatively expensive and it is sufficient to perform them once every 5 depth levels. It is necessary to keep lists of k/j-2 and k/j-1 cliques; shorter cliques may be discarded.

This algorithm was implemented in C/C++ and used to estimate graph population dynamics for R(k,j) for a number of combinations of small k and j (see table 1), using a dual-socket Intel Xeon E5-2697 v2 (24 cores at 2.7 GHz.) (Principal elements of the source code used in this analysis are available at \url{https://github.com/ekuznetsov139/ramsey}.)

Time needed for complete enumeration grows extremely rapidly with k and j. For example, there are so few R(6,5) distance colorings that they can be fully enumerated in a fraction of a second. R(6,6) takes 30 seconds with this system and algorithm, R(6,7) takes several hours, and R(6,9) would take thousands of years. Among $j \geq k \geq 5$ pairs not listed in the table, only (5,9) could be covered in less than a year. 

\begin{center}
Table 1. Longest distance colorings (full enumeration)

\begin{tabular}[h]{|c|c|c|}\hline
$(k,j)$ & Longest distance coloring & Longest known coloring ~\cite{1}~\cite{2} \\\hline

$(3,12)$ & 48 & 51 \\
$(3,13)$ & 57 & 58 \\
$(3,14)$ & 63 & 65 \\
$(3,15)$ & 72 & 72 \\
$(3,16)$ & 78 & 78 \\
$(3,17)$ & 91 & 91 \\
$(3,18)$ & 97 & 98 \\
$(3,19)$ & 105 & 105 \\
$(3,20)$ & 108 & 110 \\
%$(3,21)$ & 121 & 121 \\
$(4,5)$ & 24	& 24\\
$(4,6)$ & 33 & 35\\
$(4,7)$ & 46 & 48\\
$(4,8)$ & 52 & 57\\
$(4,9)$ & 68 & 72\\
$(4,10)$ & 91 & 91\\
$(4,11)$ & 97 & 97 \\
$(4,12)$ & 127 & 127 \\
$(4,13)$ & 136 & 137 \\
$(5,5)$ & 41	& 42\\
$(5,6)$ & 56	& 57\\
$(5,7)$ & 79	& 79\\
$(5,8)$ & 100 & 100\\
$(6,6)$ & 101 & 101\\
$(6,7)$ & 108 & 112\\
\hline
\end{tabular}
\end{center}

In regard to diagonal numbers (k=j), it takes a fraction of a second to go through all R(5,5) distance colorings. The number of colorings peaks at N=25, at 56390 (up to color reflection), and largest colorings can be seen at N=41 (there are 11 of these.) R(6,6) colorings peak at N=43, at 509235426, and the largest coloring is a cyclic coloring at N=101.

\section{Connected components}

Larger k and j present a more formidable problem. For example, for R(7,7), with $s=35$ and $d=128$, we can explore ~10 35-signatures per second. But, since there are $8.6*10^9$ possible 35-signatures, full enumeration would take ~30 years. Fortunately, it is possible to generate nearly all large R(k,j) colorings in a small fraction of time it would take to do full enumeration, utilizing connectivity of their set.

Starting with a known extensible signature, there are two easy ways to find additional signatures. One is the nearest-neighbor search: flipping one of the bits in the signature and checking if the result is extensible. The other is "relabeling". Suppose that the signature extends to a cyclic coloring of order $N \geq d$ represented by a bit mask $b$ (there's often at least one such coloring). For any $M$ that is coprime with $N$, we can relabel its vertices according to the rule $b'(x) = b((M(x+1)\mod N) - 1)$. This creates a new cyclic coloring (isomorphic to the original one) and initial bits of this coloring give us a new extensible signature. (It is not necessary to enumerate all distance colorings corresponding to each signature, since it is a relatively expensive operation. It is sufficient to construct only a single order $d$ coloring, inspect its bits, and try to idenify "nearby" cyclic colorings. In many cases there is a value of $N$ for which the symmetry constraint $b_k = b_{N-2-k}$ fails only in a few locations. We can list all candidate order $N$ cyclic colorings that differ from it in at most $n$ locations, for, say, $n=10$, and check all $2^n$ candidates.) 

For any R(k,j), if $s$ and $d$ are set sufficiently low, the set of signatures tends to be well connected: each of the two operations above connects most signatures to multiple others. Therefore, starting with a small number of signatures, we can reach a large fraction of the set. At the same time it's preferable not to set $s$/$d$ too low, since the size of the solution set grows exponentially with decreasing $d$ and time needed to check an average signature grows exponentially with decreasing $s$. (The size of the solution set depends on $s$ as well, but this dependency is weak and can be neglected.)  Empirically, the choice $d=L(k,j)-20$, $d/2-15 \leq s \leq d/2-5$, where $L(k,j)$ is the order of the largest cyclic Ramsey (k,j) graph (or, rather, its \textit{a priori} estimate), appears to work well, resulting in reasonably well connected solution sets containing $10^5$ to $10^6$ elements. 

A complication presented by this approach is that, if executed literally as described, it ends up spending most of the time reviewing a small fraction of "pathological" signatures. A pathological signature can take 1000x the time it takes to check a randomly picked signature. Fortunately, in can be observed that pathological signatures generally tend not to extend far beyond $d$ (or usually even to $d$). Therefore, this can be worked around by adding a test counter to each signature (incremented every time we try to color a link and check for new cliques), setting a scan abort threshold, and reporting "not extensible" and terminating the search if the number of tests exceeds the threshold. Optimal value of the threshold depends on values of $s$ and $d$ (all else equal, lower $s$ means that the tree rooted at each signature is larger, which means that a higher threshold is needed) and on thoroughness of the look-ahead / early termination algorithm, but, with parameters and algorithms used for this study, values in $10^4$ to $10^5$ range provided good balance between search performance and miss rate.

\section{Cyclic coloring search}

The algorithm from the previous section can find most large colorings from a small number of initial "seeds", but that still leaves the problem of finding these initial seeds. 

For some $(k,j)$, initial signatures can be generated by checking randomly picked signatures until enough seeds are generated. (The same process can provide an estimate of the size of the solution set.) However, this has limited value because it is only feasible in a few cases. For example, for R(6,8), $s=45$, $d=105$, using hardware described above, we can generate multiple signatures every minute. But already for R(6,9), $s=75$, $d=160$, yield falls below one signature per day. 

One way to solve this is to conduct a similar search, but this time on cyclic colorings rather than signatures. Define the following operations on an N-bit symmetric bitmask (order N+1 cyclic coloring):

* Bit flip:  flip bits $a$ and $N-a-1$, where $0 \leq 2*a \leq N$.

* Reflection: for $N' \neq N$, take the first $\lceil N'/2 \rceil$ bits of the bitmask and append the last $\lfloor N'/2 \rfloor$ bits of the same bitmask.

* Relabel: same as described in the previous section; reassign bits according to the rule $b'(x) = b((M(x+1)\mod (N+1) - 1)$ for $M$ coprime with $N+1$.

Each of these operations makes a new cyclic coloring. If we start with a Ramsey (k,j) cyclic coloring, some colorings reached by performing one or more of these operations will also be Ramsey (k,j) colorings. 

This suggests the following algorithm.

1. Find an initial symmetric coloring. Unlike in the previous section, there's no minimum length constraint. For (k,j) combinations considered in this article, seed colorings with orders in 100 to 150 range can be found reasonably quickly. For higher (k,j), it is possible to pick a random known (k,j-1) coloring as a starting point. 

2. Define $l_{min}$ to be the minimum coloring order we're interested in at each step. Set it initially to the order of the seed coloring.

3. Apply operations above to generate new candidates. A reasonable tradeoff between run time and search yield is achieved if we enumerate all colorings reached with up to 1 bit flip after relabel and reflection, or up to 2 bit flips without reflection. (It is not necessary to use relabel without reflection, since any coloring reached by a combination of relabel and any number of bit flips is isomorphic to a coloring that can be reached with the same number of bit flips alone.) 

4. Test all candidates. Update $l_{min}$ to $\max(l_{min}, n_{max}-8)$ where $n_{max}$ is the order of the largest coloring found during this step. Discard any colorings below $l_{min}$.

5. If no new larger colorings were found, generate a number of candidates by combining relabel, reflection and 2 to 3 bit flips at randomly picked locations, and test those as well.

6. If any colorings remain, go to step 3.

In practice, it is convenient to perform steps 3-5 in batches, processing a number of colorings (say, 100) at a time. A list of colorings is maintained, with newly discovered colorings appended to the front after each step. The value of $l_{min}$ is slowly increased if no larger colorings are found (e.g. incrementing it by 1 every time 500-1000 colorings are processed without yielding any larger colorings), ensuring that the search terminates without getting bogged down in a dead-end corner of the search space. 

It is necessary to keep a list of known and previously discovered colorings, and to check all candidates against the list to avoid going in circles.

To accelerate the process, it is possible to pre-vet candidate colorings, significantly reducing the number of colorings for which we need to execute full maximal clique search. 

* Start with the list of "submaximal" cliques (cliques with color 0 and order k-1 or color 1 and order j-1) in the coloring used as input for step 3 and 5, after relabeling but before bit flips or reflection. (Best performance is achieved if the list of cliques is generated once, before relabeling, but this presents additional technical challenges.)

* Generate a list of 1-incomplete maximal cliques in the coloring. For each submaximal clique $S$ of color 'c' and each node $V$ which is connected to the root node with a link of color '1-c', check if all links connecting $V$ with all other nodes in $S$ are also colored 'c'. If that is the case, add the clique (as a bitmask with set bits corresponding to all nodes of $S$, to node $V$, and all interconnecting links) to list $L$. Alternately, one can search for 1-incomplete cliques by executing the Appendix A algorithm directly (trying to flip each link in turn and recording maximal cliques that form) - this is generally slower but results in a "better" list.

* List values of $V$ for which there is at least one incomplete maximal clique with no nodes beyond $\lfloor N/2 \rfloor$. Skip them when picking bit flip locations during step 5. 
	
* For each candidate coloring generated during steps 3 or 5, go through $L$. It takes a single big-integer binary arithmetic operation to check if a 1-incomplete clique from $L$ is now complete in the candidate. If that is the case, the candidate can be safely discarded.

There are two additional improvements that can be employed. One is to thin the list of 1-incomplete cliques before use. The number of such cliques can at times be extremely large and all of them are not necessary to achieve adequate pre-vetting (in fact, using all of them may be counterproductive.) Usually no more than $10^3$ cliques need to be kept. Cliques with fewest numbers of bits above the position $\lfloor N/2 \rfloor$ tend to be the most valuable. (For example, during a search for diagonal R(k,k) colorings, a 1-incomplete color 'c' order $k$ clique with only a single bit above $\lfloor N/2 \rfloor$, which is colored '1-c' in the original coloring, would, on average, 'kill' around half of candidates involving reflection, because there's a 50\% chance that the missing link would change color after reflection.)

The second improvement is intelligent selection of bit flip locations during step 5. A simple but effective way is to calculate the number of triangles that would be created by flipping each bit, and pick $N/10$ to $N/5$ locations with lowest counts. 

With only slight modifications (such as setting $l_{min}=d$ and holding it fixed), this algorithm is also quite effective at enumerating large cyclic colorings. All that's needed is a number of seeds and/or a set of known colorings for R(k,j-1), and the algorithm can rapidly find all colorings connected to this starting set.

A shortcoming of this algorithm is its limited ability to connect to longer colorings from shorter colorings. Any generalization of the reflection method increases run time by orders of magnitude without a commensurate increase in yield. This appears to become an increasingly important problem beyond order 300. For low combinations (e.g. R(6,10), largest coloring order 203) there is a continuous "chain" of colorings going all the way from order 120 to the maximal order 203 coloring, with any two colorings in the chain connected via operations listed above, without more than two flips at any step. At high orders this is not always the case.

One workaround is to look for R(k,j) colorings by performing a search for R(k,j+1) colorings of the same order as described, and then checking if any of them are R(k,j) colorings. This is not always practicable because the corresponding R(k,j+1) list is often extremely long (e.g. the largest R(5,9) coloring found here is order 132, but there are over $10^6$ R(6,9) colorings at order 132 alone, and over $10^7$ at order 132 and above). Nevertheless, this approach occasionally produces colorings which couldn't be found in any other way.

Alternatively, connections may be found by employing signature-based search as described above, with slight modifications. The search may be limited to cyclic colorings of fixed order (with an explicit loop over a range of plausible order values) and executed on lowest bits of known colorings as discovered (but not on their relabelings). This allows us to see reasonable performance even with $s$ set to $d/2-30$ or lower (in essence, letting us check colorings with all possible assignments of colors to 30 highest-index links.)

\section{Coloring doubling}

Many large $(k,j)$ colorings are related to $(\lceil k/2 \rceil, j-\delta)$ colorings for $\delta \in [0;2]$. This relationship is apparent upon examination of their bitmasks (see Figure 1). Two general patterns were observed. For many $(k,j)$ colorings of order $N$, there is at least one relabeling that corresponds to a group of $(2k-1,j+1)$ colorings of order $4N$. This construction is closely related to the construction from the theorem 9 of ~\cite{6}. As in \cite{6}, the longer coloring consists of 4 copies of the order $N$ coloring, and it differs only in assignments of colors to some of the links connecting these copies. Unlike in \cite{6}, this pattern is not absolute and it only holds for some colorings.

The second pattern connects a $(k,j)$ coloring to $(2k-1,j)$, $(2k,j)$, or $(2k-1, j+2)$ colorings. (One might reasonably expect this list to include $(2k-1, j+1)$, but no notable colorings of this form were observed.) In this case, there's no firm relationship between orders, and each original coloring may correspond to a number of colorings with different orders. 

\begin{figure}[t]
\centerline{\epsfig{file=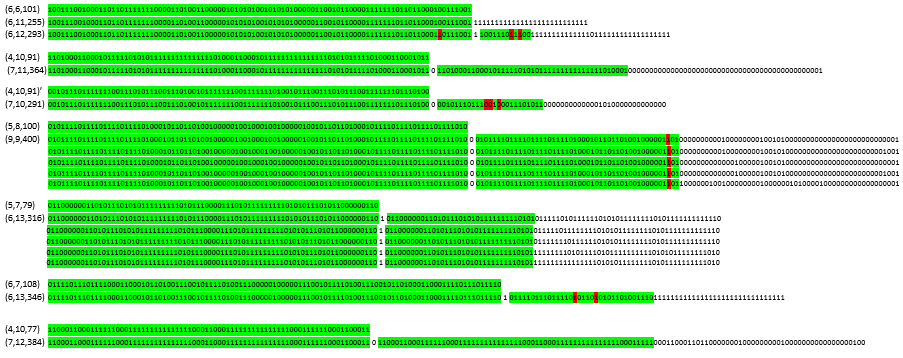,width=7in}}
\caption{Sample coloring doubling bitmasks}
\label{fig1}
\end{figure}

\section{Results}

These methods were implemented and optimized in C++ and tested at most (k,j) combinations with $j+k \leq 19$.

%Cyclic search with variable $l_{min}$ described in the previous section rapidly reaches at least as far as 65\% L(k,j) (e.g. order 145 for L(6,10)=203.) Beyond that, behavior varies considerably, depending on the combination. Combinations with both k and j odd ((7,9), (7,11)) appear to be well connected, with searches ultimately terminating at least at 85-90\% L(k,j). Other combinations exhibit "bottlenecks" near 65\% L(k,j), where connections to longer colorings become somewhat uncommon, wich severities which vary depending on parity of k and j and on the value of L(k,j). For example, (7,10) has only a mild bottleneck near order 200; (8,9)'s, in the same neighborhood, is considerably more severe; and (8,10) is so impenetrable that 60+ runs starting from randomly picked colorings (averaging 20 minutes per run) failed to produce even one coloring above order 300. (Instead, results below were obtained by running searches starting at (8,9) and (7,10) colorings.)
%Output of this process was fed into cyclic search with fixed $l_{min}$. Depth-based search (typically only on the subset of colorings with highest observed orders, due to its relatively high cost) was used to complement results of the cyclic search.

Largest observed colorings and numbers of cyclic colorings in each case are listed in tables 2-4. In most cases largest colorings were cyclic or in the neighborhood of a cyclic coloring. Colorings in table 2 were verified using third party software (Mathematica and/or Python package 'networkx'~\cite{5}.) Many other large colorings are included in the attached script. Full lists are available upon request.

In cases of $R(5,9\dots 12)$, $R(6,8\dots10)$, $R(7,7)$, and $R(7,8)$, listed counts are believed to be nearly complete. In other cases, a reasonable effort was made to generate the majority of large colorings, but there may be room for further improvement.

In several cases, largest observed colorings are larger than best previously reported and therefore represent new lower bounds on Ramsey numbers. In all remaining cases (with the exception of R(8,8)), best known lower bounds are given by non-cyclic colorings.

According to the theorem 9 of ~\cite{6}, new lower bounds on $R(5,11)$ through $R(5,14)$ imply that $R(9,12) \geq 729, R(9,13) \geq 809, R(9,14) \geq 929, R(9,15) \geq 1065$.

Coloring counts typically follow an exponential trend, with the number of colorings declining by a factor of 5 to 10 for every 5 additional nodes. Most cases exhibit systematically higher coloring counts at even orders, though this is not universal ((6,10) has the opposite pattern, and  (7,7), (8,8) and (6,12) are irregular.) Most cases, particularly ones with observed maximum orders under 250, are reasonably "well-behaved", with good adherence to the trend and fewer than 10 colorings at the maximum order. Higher cases become increasingly erratic, though it is difficult to say whether this is a genuine feature of the set of colorings or an artifact caused by deficiencies of the search algorithm.

In a few cases, there are large deviations from the exponential trend. (5,j) sets for $j \geq 10$ have spikes at orders $28*(j-10)+124$. In case of (5,12), the spike is so large that order 180 colorings account for 97\% of all observed colorings of order 177 and above, and the search produced 50 times more colorings of order 180 than colorings of orders 178 and 182 combined. Likewise, at (5,13), order 208 colorings account for 90\% of all observed colorings of order 205+. (6,12) has a large spike at order 202. (6,12) has a spike at order 235. (7,8) has spikes at 160, 168, 176 and 184. (8,8) has spikes at 193, 196, 203 and 220.

Largest observed colorings were non-cyclic for some $R(4,j)$ and cyclic in all other cases.

In case of R(7,7), current lower bound is 205, but the largest coloring constructed via these methods is only order 153. There are additional cyclic colorings of order 178 and 202, but they are "isolated". This issue is addressed in some more detail in the next section.

Colorings of order 255 and 293 were found for R(6,11) and R(6,12) respectively, using the doubling method on order101 R(6,6) coloring. This was quite unexpected because both cases were thoroughly explored with connected methods, producing maximum orders of 248 and 260 and giving no indication that longer colorings might exist. It is probable that these colorings form isolated families because their "parent" R(6,6) coloring is isolated.

In case of R(8,8), there is likewise an anomalously large isolated coloring (order 281), but there are strong indications that "regular" colorings are likely to disappear at much lower orders. The search produced a group of closely related colorings at order 220 and no other colorings larger than 206.

Higher combinations proved increasingly difficult to handle. Algorithms get slower with increasing order, though cyclic-based search performance remains tolerable even at order 400. The principal difficulty has to do with the fact that data sets grow extremely large. For example, at (7,7), it is possible to set $d=128$, take a single order 135 coloring, execute cyclic search, and, in a matter of minutes, locate the order 153 coloring as well as over 90\% of the approx. 52300 colorings at order 128 and above. On the other hand, to achieve acceptable connectivity in cases with maximum orders above 250, it is typically necessary to operate with data sets exceeding $10^6$ and at times even $10^7$ colorings. In several cases, full connectivity was not achieved at all, with observed colorings forming multiple large non-contiguous clusters.

Results for (9,9), (8,11) and (7,12) were mostly obtained via the coloring doubling method. In case of (9,9), largest colorings are order 423 and correspond to doublings of an order 132 (5,9) coloring. Current lower bound $R(9,9) \geq 565$ almost certainly can't be challenged. Largest observed (7,11) and (8,11) are order 364 and 417 and correspond to order 91 (4,10) and order 126 (6,8) colorings. In both cases, one can reasonably suspect the existence of longer cyclic colorings (order 404 and 432, related to largest (6,6) and (6,7) colorings respectively). Attempts to locate them have not been successful, but, given the nature of the doubling pattern, their existence can't be ruled out. Some other observed doublings are listed in table 5.

\begin{center}
Table 2. Largest cyclic colorings

\begin{longtable}[h]{|c|p{120mm}|}\hline
Ramsey number & Coloring \\\hline
%$(4,13)$ & CirculantGraph[136, \{1, 2, 7, 12, 14, 15, 17, 20, 24, 28, 30, 39, 40, 44, 47, 48, 55, 56, 65\}]\\
$(5,9)$ & CirculantGraph[132, \{5, 6, 9, 11, 12, 15, 16, 17, 19, 20, 27, 28, 29, 30, 31, 32, 36, 38, 39, 41, 42, 46, 53, 56, 59, 63, 66\}]\\
$(5,11)$ & CirculantGraph[182, \{4, 5, 7, 13, 21, 22, 29, 30, 31, 36, 39, 40, 43, 48, 50, 54, 61, 63, 64, 65, 66, 68, 70, 74, 75, 77, 80, 81, 82, 83, 84, 87, 88, 91\}]\\
$(5,12)$ & CirculantGraph[202, \{1, 2, 7, 9, 10, 15, 16, 19, 22, 23, 27, 28, 33, 35, 36, 41, 45, 48, 49, 51, 52, 59, 60, 62, 66, 71, 73, 76, 78, 79, 85, 86, 88, 91, 92, 97, 99, 101\}]\\
$(5,13)$ & CirculantGraph[232, \{3, 5, 6, 7, 20, 22, 23, 24, 27, 28, 29, 30, 31, 34, 35, 36, 39, 40, 41, 44, 46, 47, 48, 51, 52, 53, 55, 68, 69, 70, 72, 78, 80, 81, 82, 97, 98, \}]\\
$(5,14)$ & CirculantGraph[266, \{1, 6, 8, 9, 10, 11, 13, 39, 41, 42, 43, 44, 46, 49, 52, 55, 58, 61, 64, 65, 66, 67, 68, 69, 89, 90, 92, 93, 94, 96, 99, 102, 105, 108, 111, 114, 117, 120, 122, 123, 124, 125, 126, 127, 129, 131, 133, \}] \\
$(6,9)$ & CirculantGraph[182, \{3, 4, 6, 7, 10, 12, 15, 17, 20, 25, 26, 28, 29, 32, 38, 39, 40, 41, 42, 43, 45, 46, 51, 52, 53, 55, 56, 59, 60, 65, 66, 69, 71, 73, 79, 84, 85\}]\\
$(6,10)$ & CirculantGraph[203, \{3, 5, 6, 7, 9, 10, 11, 12, 14, 17, 19, 20, 21, 22, 24, 25, 26, 28, 34, 36, 38, 41, 42, 43, 45, 51, 53, 55, 59, 75, 79, 81, 83, 89, 91, 92, 93, 96, 97, 98, 100\}]\\
%$(6,11)$ & CirculantGraph[248, \{4, 8, 11, 12, 13, 14, 15, 17, 21, 24, 25, 26, 28, 30, 34, 40, 42, 44, 45, 46, 50, 51, 53, 55, 57, 58, 59, 60, 63, 64, 66, 68, 74, 82, 84, 87, 89, 90, 93, 95, 97, 100, 102, 116, 117, 118, 119\}] \\
$(6,12)$ & CirculantGraph[293, \{2, 3, 7, 8, 10, 11, 12, 15, 18, 26, 27, 28, 29, 32, 34, 35, 38, 39, 40, 41, 42, 44, 46, 48, 50, 51, 53, 55, 57, 59, 60, 61, 62, 63, 66, 67, 69, 72, 73, 74, 75, 83, 86, 89, 90, 91, 94, 98, 99, 103, 104, 108, 112, 113, 128\}]\\
%$(6,13)$ & CirculantGraph[333, \{2, 4, 9, 16, 18, 19, 21, 23, 25, 26, 30, 32, 33, 34, 35, 38, 40, 42, 47, 48, 49, 51, 52, 53, 58, 59, 60, 62, 65, 66, 67, 68, 70, 74, 75, 77, 79, 81, 82, 84, 87, 91, 94, 96, 98, 101, 102, 104, 109, 115, 116, 119, 125, 126, 130, 133, 147, 153\}] \\
$(7,7)$ & CirculantGraph[153, \{6, 7, 9, 10, 11, 12, 17, 19, 21, 22, 23, 25, 28, 30, 31, 32, 34, 35, 36, 37, 39, 41, 43, 44, 45, 46, 47, 48, 50, 52, 54, 58, 61, 67, 71, 73, 75\}]\\
$(7,7) (isolated)$ & CirculantGraph[202, \{3, 4, 6, 7, 11, 14, 15, 16, 20, 22, 24, 27, 29, 30, 35, 36, 39, 41, 51, 52, 53, 54, 55, 56, 57, 58, 59, 61, 63, 64, 67, 68, 69, 70, 73, 75, 76, 78, 80, 82, 83, 84, 88, 89, 91, 92, 93, 96, 99, 100, 101\}]\\
$(7,9)$ & CirculantGraph[251, \{4, 6, 7, 9, 11, 12, 13, 14, 15, 16, 17, 19, 23, 24, 25, 26, 27, 28, 29, 32, 33, 37, 39, 40, 42, 46, 47, 50, 51, 52, 53, 54, 55, 56, 60, 62, 63, 64, 65, 66, 67, 68, 70, 72, 73, 75, 83, 85, 86, 87, 88, 90, 91, 93, 95, 96, 98\}]\\
$(7,10)$ & CirculantGraph[291, \{1, 2, 4, 9, 11, 12, 13, 14, 15, 16, 17, 18, 19, 20, 21, 22, 23, 24, 28, 31, 35, 37, 41, 44, 48, 50, 54, 57, 58, 59, 60, 61, 62, 63, 64, 65, 66, 68, 70, 74, 75, 82, 83, 87, 88, 95, 96, 97, 100, 104, 105, 106, 107, 108, 112, 114, 115, 117, 125, 127, 128, 130, 138, 140, 141, 142, 143, 144, 145, 146\}]\\
$(8,10)$ & CirculantGraph[342, \{2, 6, 7, 9, 11, 12, 13, 14, 15, 18, 19, 21, 27, 28, 29, 30, 31, 32, 34, 35, 39, 40, 41, 42, 44, 46, 48, 50, 52, 53, 54, 55, 56, 59, 60, 62, 63, 64, 65, 66, 67, 73, 75, 76, 79, 80, 81, 82, 83, 85, 87, 88, 92, 96, 100, 101, 103, 105, 106, 107, 108, 112, 113, 121, 123, 125, 126, 128, 129, 133, 134, 138, 149, 154, 159, 171\}]\\
\hline
\end{longtable}
\end{center}
% 

%\begin{figure}[t]
%\centerline{\epsfig{file=figure2.png,width=7in}}
%\caption{Plot of $L_(x,y)/(x^2 y^2)$ for $(x,y)$ covered by this study}
%\label{fig1}
%\end{figure}

%
\begin{center}
Table 3. Largest cyclic coloring orders

\begin{tabular}[h]{|c|c|c|}\hline
Ramsey number & Largest observed cyclic coloring & Largest known coloring~\cite{1}~\cite{2} \\\hline
$(4,14)$ & 138 & 146 \\
$(4,15)$ & 152 & 154 \\
$(5,9)$ & \textbf{132} & 125 \\
$(5,10)$ & 146 & 148 \\
$(5,11)$ & \textbf{182} & 173 \\
$(5,12)$ & \textbf{202} & 193 \\
$(5,13)$ & \textbf{232} & 217 \\
$(5,14)$ & \textbf{266} & 241 \\
$(6,8)$ & 126 & 133 \\
$(6,9)$ & \textbf{182} & 174 \\
$(6,10)$ & \textbf{203} & 184 \\
$(6,11)$ & \textbf{255} & 252 \\
$(6,12)$ & \textbf{293} & 262 \\
$(6,13)$ & \textbf{346} & 316 \\
$(7,7)$ & 202 / 153 & 204 \\
$(7,8)$ & 202 / 192 & 215 \\
$(7,9)$ & \textbf{251} & 241 \\
$(7,10)$ & \textbf{291} & 288 \\
$(7,11)$ & 364 & 404 \\
$(7,12)$ & 407 & 416 \\
$(8,8)$ & 281 / 220 & 281 \\
$(8,9)$ & \textbf{328} & 316 \\
$(8,10)$ & \textbf{342} & \\
$(8,11)$ & 421 & 432 \\
$(9,9)$ & 562 / 423 & 564 \\
$(9,10)$ & 490 & 580 \\
\hline
\end{tabular}
\end{center}

Improved lower bounds are in boldface.

Note: for (7,7), (7,8), (8,8), and (9,9), the first quoted number is for the isolated coloring, the second number is for the connected coloring.

\begin{center}
Table 4. Numbers of observed cyclic colorings
\begin{longtable}[h]{|c|c|c|c|c|c|c|c|}
\hline
N & $(6,8)$ & $(5,9)$ & $(5,10)$ & $(7,7)$ & $(6,9)$ & $(5,11)$ & $(7,8)$ \\
\hline
110 & 1240  & 7043  &  &  &  &  & \\
111 & 807  & 972  &  &  &  &  & \\
112 & 515  & 1691  &  &  &  &  & \\
113 & 376  & 566  &  &  &  &  & \\
114 & 114  & 699  &  &  &  &  & \\
115 & 224  & 268  &  &  &  &  & \\
116 & 54  & 396  &  &  &  &  & \\
117 & 23  & 207  &  &  &  &  & \\
118 & 2  & 84  &  &  &  &  & \\
119 & 17  & 24  &  &  &  &  & \\
\hline
120 & 0  & 91  &  &  &  &  & \\
121 & 1  & 4  &  &  &  &  & \\
122 & 8  & 46  &  &  &  &  & \\
123 & 3  & 17  &  &  &  &  & \\
124 & 2  & 9  &  &  &  &  & \\
125 & 13  & 0  &  &  &  &  & \\
126 & 2  & 2  &  &  &  &  & \\
127 &  & 1  &  &  &  &  & \\
128 &  & 3  &  &  &  &  & \\
129 &  & 0  &  &  &  &  & \\
\hline
130 &  & 0  & 6076  & 8129  &  &  & \\
131 &  & 0  & 597  & 4174  &  &  & \\
132 &  & 1  & 5797  & 4303  &  &  & \\
133 &  &  & 255  & 4630  &  &  & \\
134 &  &  & 1960  & 3605  &  &  & \\
135 &  &  & 67  & 1571  &  &  & \\
136 &  &  & 563  & 1419  &  &  & \\
137 &  &  & 99  & 1073  &  &  & \\
138 &  &  & 274  & 408  &  &  & \\
139 &  &  & 12  & 1044  &  &  & \\
\hline
140 &  &  & 499  & 365  &  &  & \\
141 &  &  & 0  & 322  &  &  & \\
142 &  &  & 46  & 390  &  &  & \\
143 &  &  & 2  & 97  &  &  & \\
144 &  &  & 2  & 86  &  &  & \\
145 &  &  & 0  & 19  &  &  & \\
146 &  &  & 1  & 53  &  &  & \\
147 &  &  &  & 67  &  &  & \\
148 &  &  &  & 6  &  &  & \\
149 &  &  &  & 10  &  &  & \\
\hline
150 &  &  &  & 7  &  &  & \\
151 &  &  &  & 3  &  &  & \\
152 &  &  &  & 3  &  &  & \\
153 &  &  &  & 1  &  &  & \\
154 &  &  &  & 0  &  &  & \\
155 &  &  &  & 0  &  &  & \\
156 &  &  &  & 0  &  &  & \\
157 &  &  &  & 0  &  &  & \\
158 &  &  &  & 0  &  &  & \\
159 &  &  &  & 0  &  &  & \\
\hline
160 &  &  &  & 0  & 14167  & 6094  & \\
161 &  &  &  & 0  & 527  & 1321  & \\
162 &  &  &  & 0  & 2068  & 1916  & \\
163 &  &  &  & 0  & 581  & 492  & \\
164 &  &  &  & 0  & 2081  & 628  & \\
165 &  &  &  & 0  & 285  & 507  & \\
166 &  &  &  & 0  & 640  & 520  & \\
167 &  &  &  & 0  & 89  & 94  & \\
168 &  &  &  & 0  & 488  & 53  & \\
169 &  &  &  & 0  & 55  & 42  & \\
\hline
170 &  &  &  & 0  & 444  & 57  & 2722 \\
171 &  &  &  & 0  & 27  & 9  & 186 \\
172 &  &  &  & 0  & 587  & 30  & 378 \\
173 &  &  &  & 0  & 15  & 3  & 153 \\
174 &  &  &  & 0  & 189  & 18  & 1896 \\
175 &  &  &  & 0  & 2  & 0  & 19 \\
176 &  &  &  & 0  & 13  & 3  & 3358 \\
177 &  &  &  & 0  & 0  & 0  & 11 \\
178 &  &  &  & 1  & 1  & 0  & 67 \\
179 &  &  &  & 0  & 0  & 0  & 1 \\
\hline
N & $(6,8)$ & $(5,9)$ & $(5,10)$ & $(7,7)$ & $(6,9)$ & $(5,11)$ & $(7,8)$ \\
\hline
\end{longtable}

\begin{longtable} [h]{|c|c|c|c|c|c|c|c|}\hline
N & $(7,7)$ & $(7,8)$ & $(5,12)$ & $(6,10)$ & $(8,8)$ \\
\hline
180 & 0  & 227  &  & 415  & \\
181 & 0  & 0  & 7366  & 738  & \\
182 & 0  & 19  & 43167  & 197  & \\
183 & 0  & 8  & 4219  & 3896  & \\
184 & 0  & 705  & 7565  & 351  & \\
185 & 0  & 0  & 1241  & 47  & \\
186 & 0  & 6  & 14696  & 168  & \\
187 & 0  & 0  & 810  & 1  & \\
188 & 0  & 0  & 2209  & 11  & \\
189 & 0  & 0  & 139  & 531  & \\
\hline
190 & 0  & 0  & 1274  & 3  & \\
191 & 0  & 0  & 41  & 98  & \\
192 & 0  & 4  & 171  & 1  & \\
193 & 0  & 0  & 116  & 164  & \\
194 & 0  & 0  & 555  & 0  & \\
195 & 0  & 0  & 1  & 129  & 933 \\
196 & 0  & 0  & 21  & 0  & 5177 \\
197 & 0  & 0  & 2  & 2  & 48 \\
198 & 0  & 0  & 102  & 0  & 42 \\
199 & 0  & 0  & 0  & 4  & 0 \\
\hline
200 & 0  & 0  & 0  & 0  & 16 \\
201 & 0  & 0  & 0  & 1  & 1 \\
202 & 1  & 2  & 8  & 0  & 3 \\
203 &  &  &  & 1  & 1453 \\
204 &  &  &  &  & 5 \\
205 &  &  &  &  & 0 \\
206 &  &  &  &  & 1 \\
207 &  &  &  &  & 0 \\
208 &  &  &  &  & 0 \\
209 &  &  &  &  & 0 \\
\hline
210 &  &  &  &  & 0 \\
211 &  &  &  &  & 0 \\
212 &  &  &  &  & 0 \\
213 &  &  &  &  & 0 \\
214 &  &  &  &  & 0 \\
\hline
N & $(7,7)$ & $(7,8)$ & $(5,12)$ & $(6,10)$ & $(8,8)$ \\
\hline
\end{longtable}

\begin{longtable} [h]{|c|c|c|c|c|c|c|c|}\hline
N & $(8,8)$ & $(5,13)$ & $(6,11)$ & $(7,9)$ & $(6,12)$ & $(5,14)$ \\
\hline
215 & 0  & 13783  &   &  &  & \\
216 & 0  & 40736  &   &  &  & \\
217 & 0  & 7893  &   &  &  & \\
218 & 0  & 6199  &   &  &  & \\
219 & 0  & 3915  &   &  &  & \\
\hline
220 & 130  & 2270  & 6379  &  &  & \\
221 &  & 1183  & 1694  &  &  & \\
222 &  & 3516  & 257  &  &  & \\
223 &  & 786  & 135  &  &  & \\
224 &  & 473  & 3075  &  &  & \\
225 &  & 182  & 5  & 83669 &  & \\
226 &  & 198  & 5491  & 98477 &  & \\
227 &  & 68  & 61  & 23198  &  & \\
228 &  & 137  & 73  & 126830  &  & \\
229 &  & 7  & 40  & 18185  &  & \\
\hline
230 &  & 34  & 95  & 48831 &  & \\
231 &  & 3  & 41  & 5319  &  & \\
232 &  & 10  & 83  & 12247  &  & \\
233 &  &  & 7  & 13166  &  & \\
234 &  &  & 10  & 15888 &  & \\
235 &  &  & 2  & 1584  &   & \\
236 &  &  & 19  & 2593  &   & \\
237 &  &  & 15  & 2825  &   & \\
238 &  &  & 12  & 1266  &   & \\
239 &  &  & 0  & 444  &   & \\
\hline
240 &  &  & 5  & 464  &   & \\
241 &  &  & 0  & 574  &   & \\
242 &  &  & 17  & 223  &   & \\
243 &  &  & 0  & 217  &   & \\
244 &  &  & 0  & 20  &   & \\
245 &  &  & 0  & 33  &   & 7687 \\
246 &  &  & 8  & 1  &  & 10582 \\
247 &  &  & 0  & 0  &   & 1742 \\
248 &  &  & 3  & 0  &   & 2859 \\
249 &  &  & 0 & 0  &   & 622 \\
\hline
250 &  &  & 4  & 0  &   & 728 \\
251 &  &  & 0 & 1  & & 79 \\
252 &  &  & 0 &  & & 308 \\
253 &  &  & 0 &  & & 38 \\
254 &  &  & 0 &  &   & 515 \\
255 &  &  & 10 &  &   & 25 \\
256 &  &  &  &  &   & 28 \\
257 &  &  &  &  &   & 1 \\
258 &  &  &  &  &   & 600 \\
259 &  &  &  &  &   & 0 \\
\hline
260 &  &  &  &  & 4  & 2 \\
261 &  &  &  &  & 797 & 0 \\
262 &  &  &  &  & 445 & 0 \\
263 &  &  &  &  & 286 & 4 \\
264 &  &  &  &  & 86 & 0 \\
265 &  &  &  &  & 159 & 0 \\
266 &  &  &  &  & 0 & 32 \\
267 &  &  &  &  & 0 &  \\
268 &  &  &  &  & 58 & \\
269 &  &  &  &  & 218 & \\
\hline
N & $(8,8)$ & $(5,13)$ & $(6,11)$ & $(7,9)$ & $(6,12)$ & $(5,14)$ \\
\hline
\end{longtable}

\begin{longtable} [h]{|c|c|c|c|c|c|c|c|}\hline
N & $(7,10)$ & $(6,12)$  \\
\hline
270 & 6572  &  0 \\
271 & 49934 &  250 \\
272 & 119670  &  2 \\
273 & 0  &  0 \\
274 & 378931  &  13 \\
275 & 14602  &  78 \\
276 & 123 &   19 \\
277 & 13315  &  94 \\
278 & 3137  &  0 \\
279 & 0  &  0 \\
\hline
280 & 350 &  0 \\
281 & 4361  &  0 \\
282 & 1228  &  0 \\
283 & 90  &  0 \\
284 & 0  &  0 \\
285 & 38  &  7 \\
286 & 0  & 0 \\
287 & 10  &  0 \\
288 & 35 &  0 \\
289 & 336  &  0 \\
\hline
290 & 0  &  0\\
291 & 141  &  0 \\
292 &  &  0 \\
293 &  &  1 \\
\hline
N & $(7,10)$ & $(6,12)$  \\
\hline
\end{longtable}

\begin{longtable} [h]{|c|c|c|c|c|}\hline

N & $(6,13)$ & $(8,9)$ & $(8,10)$ & $(7,11)$ \\
\hline
310  & 10814  & 293  &  & \\
311  & 989  & 167  &  & \\
312  & 3727  & 8327  &  & \\
313  & 832  & 0  &  & \\
314  & 2845  & 11898  &  & \\
315  & 5  & 91  &  & \\
316  & 870  & 58  &  & \\
317  & 992  & 0  &  & \\
318  & 96  & 54  &  & \\
319  & 4401  & 12  &  & \\
\hline
320 & 14  & 0  & 19638  & \\
321 & 312  & 26  & 77825  & \\
322 & 0  & 0  & 4406  & \\
323 & 67  & 0  & 4454  & \\
324 & 2  & 0  & 5848 & \\
325 & 0  & 0  & 14  & 97 \\
326 & 77  & 0  & 561  & 2481 \\
327 & 2  & 0  & 0  & 4379 \\
328 & 15  & 13  & 3683  & 1100 \\
329 & 0  &  & 0  & 504 \\
\hline
330 & 0  &  & 41  & 22 \\
331 & 0  &  & 0  &  17 \\
332 & 0  &  & 16  & 14 \\
333 & 7  &  & 0  & 71 \\
334 &  &  & 0  & 85 \\
335 &  &  & 0  & 1 \\
336 &  &  & 1  & 49 \\
337 &  &  & 172  & 2 \\
338 &  &  & 0  & 202 \\
339 &  &  & 0  & 0 \\
\hline
340 &  &  & 0  & 1 \\
341 &  &  & 0  & 0 \\
342 &  &  & 14  & 0 \\
343 &  &  &   & 0 \\
344 &  &  &   & 0 \\
345 &  &  &   & 0 \\
346 &  &  &   & 2 \\
$(\dots)$  &  &  &   & 0 \\
364 &  &  &   & 8866 \\
\hline
N & $(6,13)$ & $(8,9)$ & $(8,10)$ & $(7,11)$ \\
\hline
\end{longtable}

\begin{longtable} [h]{|c|c|c|c|}\hline
N & $(7,12)$ & $(8,11)$ & $(9,9)$ \\
\hline
400 & 0  & & 21564 \\
401 & 0  & 6  & 115 \\
402 & 0  & 21  & 240 \\
403 & 58  & 3722  & 14138 \\
404 & 0  & 425192 & 8 \\
405 & 0  & 605 & 301 \\
406 & 0  & 0  & 790 \\
407 & 988 & 0  & 0 \\
408 &  & 698593 & 36007 \\
409 &  & 4003  & 2407 \\
\hline
410 &  & 44884 & 33531 \\
411 &  & 0  & 2 \\
412 &  & 22151  & 12 \\
413 &  & 0  & 0 \\
414 &  & 0  & 0 \\
415 &  & 0  & 0 \\
416 &  & 6  & 0 \\
417 &  & 2102  & 0 \\
418 &  & 0  & 0 \\
419 &  & 0  & 4 \\
\hline
420 &  & 35  & 0 \\
421 &  & 2  & 0 \\
422 &  &  & 0 \\
423 &  &  & 19 \\
\hline
\end{longtable}

Table 5. Doubling connections
\begin{longtable} [h]{|c|c|c|c|c|c|c|}\hline
k,j & New & $\lceil k/2 \rceil, j$ & $k, \lceil j/2 \rceil$ & 4x repeat  & Others \\
\hline
6,11 & & & $(6,6): 101 \rightarrow 255$ & & \\
6,12 & & & $(6,6): 101 \rightarrow 293$ & & \\
6,13 & & & $(6,7): 108 \rightarrow 346$ & $(5,7): 79 \rightarrow 316$ & \\
6,14 & & & $(6,7): 108 \rightarrow 356$ & & \\
7,7 & & $(4,7): 46 \rightarrow 153$ & & & \\
7,8 & 192 & (4,8): $51 \rightarrow 184$ & & & \\
7,9 & & $(4,9): 68 \rightarrow 241$ & & & $(4,7): 46 \rightarrow 234$ \\
7,10 & & $(4,10): 91 \rightarrow 291$ & & $(4,9): 68 \rightarrow 272$ & $(4,8): 51 \rightarrow 268$ \\
7,11 & & $(4,11): 95 \rightarrow 346$ & $(7,6): 108 \rightarrow 335$ & $(4,10): 91 \rightarrow 364$ & $(4,9): 68 \rightarrow 320$ \\
7,12 & & $(4,12): 127 \rightarrow 407$ & $(7,6): 108 \rightarrow 357$ & $(4,11): 95 \rightarrow 380$ & $(4,10): 91 \rightarrow 389$ \\
8,8 & 220 & & & & \\
8,9 & & & $(8,5): 93 \rightarrow 328$ & $(7,5): 79 \rightarrow 316$ & \\
8,10 & & $(4,10): 91 \rightarrow 312$ & & & \\
8,11 & & $(4,11): 95 \rightarrow 364$ & $(8,6): 125 \rightarrow 421$ & & \\
9,9 & & $(5,9): 132 \rightarrow 423$ & & $(5,8): 100 \rightarrow 400$ & \\
9,10 & & $(5,10): 142 \rightarrow 490$ & & & $(5,8): 100 \rightarrow 406$ \\
\hline
\end{longtable}

\end{center}

\section{Isolated colorings}

In all cases, we can't be sure that true maximal colorings are not isolated (with no large "nearby" colorings). We can cover connected components of the solution set but we can't find isolated colorings without full enumeration. One could reasonably expect that most maximal colorings are inside connected components, but this is not always the case. The ones most easily missed are colorings that may be called "degenerate". Though a general cyclic coloring would appear in the solution set as many distinct versions and therefore many distinct signatures (see "relabeling" above), making it relatively hard to miss (all we need is for one of those relabelings to be in a cluster), for some colorings, different relabelings are not just isomorphic but equal. (One example is the Paley graph of order 101: all its relabelings are identical, up to color reflection.) To address this limitation, we can directly construct and review colorings with fewest distinct relabelings.

The set of relabelings of a coloring of order N is isomorphic to the multiplicative group of integers modulo N. Therefore, to construct these colorings, we need to examine the structure of this group.

For prime N, the situation is easiest. The group is cyclic and has N-1 elements. There is a family of degenerate colorings for each subgroup of $C_{N-1}$ (in other words, for each factor of $N-1$.) We can construct a fully degenerate coloring, with only one unique relabeling, by setting $C(q^k mod N) \equiv C(1,1+(q^k) mod N) = (1+(-1)^k)/2$ where $q$ is any generator of the group. (It is easy to see that, if N is a Pythagorean prime, this coloring is none other than the Paley coloring of order N.) If $N-1$ is divisible by 3, there are 6 nontrivial colorings with 3 distinct relabelings each, distinguished by colors of $q^k, 1\leq k \leq 3$ (coloring $C(q)=C(q^2)=C(q^3)=0$ is trivial, and coloring $C(q)=C(q^3)=1, C(q^2)=0$ is fully degenerate.)

For $N=2p$ with $p$ prime, the group is also cyclic and links of the coloring form three orbits: links with numbers coprime with $N$; links with even numbers; and link $p$. Link $p$ is fixed under group action. Suppose that $q$ is a generator of the group. There are 16 colorings with either 1 or 2 distinct relabelings, one for each combination of colors of links $2$, $q$, $2q$, and $p$. (Link 1 is held fixed. Values of $q$ and $2q$ indicate whether links 1 and 2 are held constant or flipped under action of the generator.)

More generally, take any k-element subgroup $G$ of $\mathbb{Z}/(N)$, and construct a quotient $Q = (\mathbb{Z}/(N)) / G$ that contains $N-1$ (in essence, breaking relabeling symmetry under $G$.) Action of $Q$ induces a number of orbits among links. Pick an arbitrary 'initial' link in each orbit. Suppose there are $a$ orbits and $b$ generators in $Q$. We can construct $2^{a+b-1}$ partially degenerate colorings, for each combination of value assigned to each 'initial' link and to each action of the generator. 

For $100<N<250$ and $G$ generated by each element of $\mathbb{Z}/(N)$, colorings with up to 22 orbits and fixed action of generators, and colorings with up to $a+b\leq 20$, were constructed. In addition, first two basic classes above were constructed for $N<1000$. For off-diagonal Ramsey numbers, no notable colorings were found (largest observed colorings were: for R(5,10), N=140; for R(6,8), N=116; for R(6,9), N=156; for R(6,10), N=174.) However, for R(7,7), $N=2p$ class yielded a cyclic coloring of order 202. It is degenerate (with only 2 distinct relabelings differing by a single bit) as well as deeply isolated (with no neighbor signatures extending even to 100), making it virtually impossible to find via nearest-neighbor search. (The same coloring also works as the largest cyclic coloring for R(7,8).) This coloring is isomorphic to a block-  coloring with adjacency matrix $(\frac {A \bar{A}} {\bar{A} A})$ where $A$ is the Paley coloring of order 101, and, therefore, is a subset of the order 204 coloring found by Shearer (1986)~\cite{3} / Mathon (1987) which is the largest currently known R(7,7) coloring.~\cite{7} For R(8,8), R(9,9), and R(10,10), cyclic colorings of order 281, 562, and 797 respectively were produced. These are likewise either Paley colorings (for (8,8) and (10,10)) or Shearer/Mathon subsets (for (9,9).)

\section{Appendix A}

Objective: given a distance coloring and a new link 'b' (from vertex 0 to vertex 'b') with color 'c', compile a list of monochromatic cliques containing this link.

The most straightforward approach would be to keep a list of all known cliques and to check if the new link extends any of them. However, this is costly memory-wise and, for large k, most cliques are not extensible by any given link (e.g., a k=6 clique has, on average, only a 1/32 chance of being extensible to k=7 with a link of the same color.) The following approach avoids this at the cost of some bit arithmetic, which can be accelerated using native x86 CPU instructions. It is in a sense a variation of the classic Bron-Kerbosch maximal clique algorithm~\cite{4}, adapted to take advantage of symmetries of graphs in question.

The algorithm employs a type 'bigint' with several methods (len(), set(), etc) whose functions should be self-evident.

1. Start with a bitmask 'x' containing bits for each link that is colored with color 'c'. Create a copy  with link 'b' set:

\lstdefinestyle{customc}{
  belowcaptionskip=1\baselineskip,
  breaklines=true,
%  frame=L,
  xleftmargin=\parindent,
  language=C,
  showstringspaces=false,
%  basicstyle=\footnotesize\ttfamily,
  keywordstyle=\bfseries\color{green!40!black},
  commentstyle=\itshape\color{purple!40!black},
  identifierstyle=\color{blue},
  columns=fullflexible,
  stringstyle=\color{orange},
}

\begin{lstlisting}[style=customc]
bigint x; 
bigint y = x;
y.set(b-1);
\end{lstlisting}

2. Define an operation 
\begin{lstlisting}[style=customc]
bigint invert(bigint x, int n)
{
   bigint y = 0;
   for(int i=0; i<x.len(); i++)
   {
        if(x.bit(i))
	{
		if(n-i-1>=0)
			y.set(n-i-1);
		if(n+i+1<y.max_len())
			y.set(n+i+1);
	}
   }
}
\end{lstlisting}

(for a 'bigint' with fixed maximum length, this can be optimized, eliminating an explicit loop over all bits in favor of shifts and bitwise or's / and's)

3. Define a stack of bigint variables. 
\begin{lstlisting}[style=customc]
int depth=0;
bigint stack[MAX_K]; 
\end{lstlisting}

4. Set the first entry in the stack:
\begin{lstlisting}[style=customc]
stack[depth] = x & invert(y, b-1);
\end{lstlisting}

5. At this point, every bit 'n' set in stack[0] corresponds to a 3-clique (0,n+1,b). Scan through bits using x86 CPU instructions bsr/bsf (exposed by under various names by different C compilers, e.g. as $\_\_builtin\_ctzll$ / $\_\_builtin\_clzll$ by gcc). For each found bit, report a clique if necessary and move on to the next step:

6.  
\begin{lstlisting}[style=customc]
    stack[depth].unset(n);
    stack[depth+1] = stack[depth] & invert(y, n);
\end{lstlisting}
    Check that the newly created bigint is not zero. If it is not, increment 'depth' and continue the process in steps 4-6 recursively. If it is, go back down one level.
	
It is preferable to compute invert() values for all bits set in y in advance and simply to load them from memory instead of executing invert() every time. 

If we're only interested in cliques which have at least $L$ nodes for some value of $L$, there are additional optimizations that can be employed. First, we can check the number of bits in $stack[depth+1]$ (also accelerated with x86 CPU instruction popcnt) at the end of step 6. If there aren't enough bits left in $y$ to construct such a clique, we can bypass the recursion and go back to the previous level.

In addition, it's possible to utilize symmetries of the underlying coloring to reduce the number of operations. 

In a distance coloring, a monochromatic clique with elements $(0, y_1, y_2, \dots y_{n-1}, y_n)$, where $0 < y_1 < y_2 < \dots <y_n$, always has a counterpart, a monochromatic clique $(0, y_n-y_{n-1}, y_n-y_{n-2}, \dots, y_n)$. Only one of the two needs to be constructed via this process (the other can be imputed). Therefore, if $L>2$, bits are always set in increasing order (or if any coloring is going to be checked at some later point by doing an ordered rebuild), step 5 scan in performed in decreasing bit order, and, at any point, we encounter the situation where $depth \leq \lfloor (L-3)/2 \rfloor$ and the largest remaining set bit in $stack[depth]$ is less than $y_n/2$, we can likewise stop the recursion and go back one level immediately.

If the coloring is cyclic, rotation symmetry means that, for each monochromatic clique $(0, y_1, y_2, \dots y_{n-1}, y_n)$, there are monochromatic cliques $(0, y_{k+1}-y_{k}, y_{k+2}-y_{k}, \dots, y_n-y_k, N-y_k, N+y_1-y_k, \dots)$ for $1 \leq k \leq n-1$, and $(0, N-y_n, N+y_1-y_n, \dots, N+y_{n-1}-y_n)$. One way to utilize this symmetry is to note that we can always rotate node labels to put the widest gap between any two adjacent node indices in the clique at the end. Therefore we can restrict the search to cliques that satisfy $y_1 \leq N-y_n$, $y_k-y_{k-1} \leq N-y_n$ for $2 \leq k \leq n$ (and, therefore, $n(N-y_n)\geq N$.)

\section{Appendix B}
Objective: given an incompletely colored distance graph, a new link $b$ with color $c$, and a number $k$ representing maximum allowed color $c$ clique size, compile a list of link-color assignments implied by this assignment.

This is straightforward with a double-loop over all monochromatic order $k-1$ cliques containing $b$ and all uncolored links in the graph. The following method is somewhat more efficient because it only requires the inner loop to perform $k-1$ iterations.

1. Execute appendix A algorithm to construct the list $v$ monochromatic order $k-1$ cliques containing $b$.

2. Define a bigint variable $M$. For each yet uncolored link in the graph, set the corresponding bit in $M$. 

3. Define a bigint variable $m$. For each color $c$ link in the graph, set the corresponding bit in $m$.

4. Construct a list $shifted\_masks$ of inverted bitmasks by executing 'invert' from appendix A on $m$ for values of $n$ corresponding to all links colored to $c$.

5. Iterate over all cliques from step 1:

\begin{lstlisting}[style=customc]
bigint new_mask;
new_mask.clear();
for(int i=0; i<v.size(); i++)
{
  bigint y = v[i];
  bigint cand = M;
  while(!y.zero())
  {
      int pos = y.trailing_bit();
      cand &= shifted_masks[pos]];
      if(cand.zero())
        break;
     y.unset(pos);
   }
  new_mask|=cand;
  M &= ~cand;
}
\end{lstlisting}

At the end of the loop, bits set in $new\_mask$ correspond to links that have to be colored $1-c$ to prevent formation of order $k$ monochromatic cliques.

\section{Appendix C}

Objective: given an incomplete coloring, numbers $d$ and k, attempt to determine if the coloring is extensible to a R(k,k) coloring of order $d$, without actually enumerating all its descendants.

1. Compile a list of 2-incomplete and 3-incomplete k-cliques. This list does not have to be comprehensive; though the comprehensive list would cut down on the search space by terminating branches as early as possible, actually constructing one for every invocation of this algorithm is substantially more expensive than simply enumerating descendants in conjunction with some fast generation method. Record colors of cliques and missing links.

In the code used in this study, this was done as follows.

* Maintain a list of complete (k-2)-cliques for each coloring. (This is relatively cheap during depth-first search.) 

* For each clique, iterate over yet-uncolored links under $d$. For a clique $(0,a_1, a_2, ... a_{k-3})$ and an uncolored link m, check if all links $|a_1-m|-1$, $|a_2-m|-1$, etc. are colored the same color as the clique. Make a list of links satisfying the requirement.

* For every pair $m_1$, $m_2$  belonging to the same clique, if $|m_1-m_2|-1$ is colored the same color 'c' as the clique, we have a 2-incomplete $(0, a_1, a_2, \dots , a_{k-3}, m_1, m_2)$, and record ${c, m_1, m_2}$.

The rest of the algorithm assumes that only 2-incomplete cliques are used; generalization to 3-incompletes is relatively straightforward.

2. Define an array of bits $v[d][4][d]$ (convenient to group as $d*4$ bigints). Initially set all of them to 0 except for v[i][0][i] and v[i][3][i], $0\leq i<d$. (v[i][0] and v[i][1] correspond to valid extensions of the coloring with bit i set to 0. v[i][2] and v[i][3] correspond to extensions with i set to 1.)

3. Go through the list of incomplete cliques. For each 2-incomplete ${c, m_1, m_2}$, set bits $v[m_1][1+c][m_2]$ and $v[m_2][1+c][m_1]$.

4. Execute a double-loop over all yet uncolored links below $d$:

\begin{lstlisting}[style=customc]
bigint mask0, mask1; // currently set bits

for(int p=0; p<d_min; p++)
{
   if(mask0.set(p) || mask1.set(p))
      continue;
   for(int i=0; i<d_min; i++)
   {
       if(i==p)
		continue;
     if(v[p][0].bit(i))	
     {
       v[p][0] |= v[i][0];
       v[p][1] |= v[i][1];
     }
     if(v[p][1].bit(i))	
     {
       v[p][0] |= v[i][2];
       v[p][1] |= v[i][3];
     }
     if(v[p][2].bit(i))	
     {
       v[p][2] |= v[i][0];
       v[p][3] |= v[i][1];
     }
     if(v[p][3].bit(i))	
     {
       v[p][2] |= v[i][2];
       v[p][3] |= v[i][3];
     }
   }
}
\end{lstlisting}

5. If, for any p, any bit is set simultaneously in v[p][0] and v[p][1], link p can't be colored to 0. Color it to 1 and recursively apply all known rules (for any 2-incomplete {1, p, q}, color q to 0, etc.)
If any bit is set in v[p][2] and v[p][3], color link p to 0 and apply rules. If this process results in an attempt to color any link in both colors at once, the coloring is not extensible to $d$. Abort the process.

6. For all bits set during step 5, apply rules to the array v. E.g. if p is colored to 0, execute
\begin{lstlisting}[style=customc]
for(int i=0; i<d_min; i++)
{
	if(i==p)
		continue;
	v[i][0] |= v[p][0];
	v[i][1] |= v[p][1];
	v[i][2] |= v[p][0];
	v[i][3] |= v[p][1];
}
\end{lstlisting}
7. Repeat the process until no further changes occur in v.

\end{document}